\documentclass[structabstract]{aa} 
\usepackage{graphicx}
\usepackage{txfonts}
\usepackage{natbib}
\usepackage{color}
\begin{document}

\title{ MOND prediction of a~new giant shell in the elliptical galaxy NGC\,3923}
% \subtitle{ }

\author{M. B\'{i}lek\inst{1}\fnmsep\inst{2}
\and
K. Barto\v{s}kov\'{a}\inst{1}\fnmsep\inst{3}
\and
I. Ebrov\'{a}\inst{1}\fnmsep\inst{4}
\and
B. Jungwiert\inst{1}\fnmsep\inst{2}}

\institute{Astronomical Institute, Academy of Sciences of the Czech Republic, Bo\v{c}n\'{i} II 1401/1a, CZ-141\,00 Prague, Czech Republic\\
\email{bilek@asu.cas.cz}
\and
Faculty of Mathematics and Physics, Charles University in Prague, Ke~Karlovu 3, CZ-121 16 Prague, Czech Republic
\and 
Department of Theoretical Physics and Astrophysics, Faculty of Science, Masaryk University, Kotl\'a\v{r}sk\'a 2, CZ-611\,37 Brno, Czech Republic
\and
Institute of Physics, Academy of Sciences of the Czech Republic, Na~Slovance 
1999/2 CZ-182\,21 Prague, Czech Republic
}

\date{Received April 03, 2014; accepted April 28, 2014}

% 5 {} token are mandatory
%%%%%%%%%%%%%%%%%%%%%%%%%%%%%%%%%%%%%%%%%%%%%%%%%%%%%%%%%%%%%%%%% 
\abstract
%%%%%%%%%%%%%%%%%%%%%%%%%%%%%%%%%%%%%%%%%%%%%%%%%%%%%%%%%%%
%%%%%%%%%%%%%%%%%%%%%%%%%%%%%%%%%%%%%%%%%%%%%%%%%%%%%%%%%%%
% context heading (optional)
{Stellar shells, which form axially symmetric systems of arcs in some elliptical galaxies, are most likely remnants of radial minor mergers. They are observed up a radius of $\sim$100\,kpc. The stars in them oscillate in radial orbits. The radius of a shell depends on  the free-fall time at the position of the shell and on the time since the merger. We previously verified the consistency of shell radii in the elliptical galaxy NGC\,3923 with its most probable MOND potential. Our results implied that an as~yet undiscovered shell exists at the outskirts of the galaxy.}
%%%%%%%%%%%%%%%%%%%%%%%%%%%%%%%%%%%%%%%%%%%%%%%%%%%%%%%%%%%
%%%%%%%%%%%%%%%%%%%%%%%%%%%%%%%%%%%%%%%%%%%%%%%%%%%%%%%%%%%
% aims heading (mandatory)
{We here extend our study by assuming more general models for the gravitational potential to verify the prediction of the new shell and to estimate its position.}
%%%%%%%%%%%%%%%%%%%%%%%%%%%%%%%%%%%%%%%%%%%%%%%%%%%%%%%%%%%
%%%%%%%%%%%%%%%%%%%%%%%%%%%%%%%%%%%%%%%%%%%%%%%%%%%%%%%%%%%
% methods heading (mandatory)
{We tested the consistency of the shell radial distribution observed in NGC\,3923 with a wide variety of MOND potentials of the galaxy. The potentials differed in the mass-to-light ratio and in distance to the galaxy. We considered different MOND interpolation functions, values of the acceleration constant $a_0$, and density profiles of the galaxy. We verified the functionality of our code on a Newtonian self-consistent simulation of the formation of a~shell galaxy.}
%%%%%%%%%%%%%%%%%%%%%%%%%%%%%%%%%%%%%%%%%%%%%%%%%%%%%%%%%%%
%%%%%%%%%%%%%%%%%%%%%%%%%%%%%%%%%%%%%%%%%%%%%%%%%%%%%%%%%%%
% results heading (mandatory)
{Our method reliably predicts that exactly one new outermost shell exists at a galactocentric radius of about 1900$^{\prime\prime}$ ($\sim$210\,kpc) on the southwestern side of the galaxy. Its estimated surface brightness is about 28\,mag\,arcsec$^{-2}$ in $B$ -- a value accessible by current instruments. This prediction enables a  rare test of MOND in an elliptical down to an acceleration of $a_0/10$. The predictive power of our method is verified by reconstructing the position of the largest  known  shell from the distribution of the remaining shells. }
% conclusions heading (optional), leave it empty if necessary 
{}


\maketitle
%%%%%%%%%%%%%%%%%%%%%%%%%%%%%%%%%%%%%%%%%%%%%%%%%%%%%%%%%%%%%%%%%

\section{Introduction} \label{sec:intro}
Modified Newtonian dynamics (MOND) and its implications has been successfully tested in all types of disk and dwarf galaxies, and its numerous aspects have been even verified in interacting galaxies \citep{famaey12}.
However, tests of MOND in ellipticals are still rare. Ellipticals typically lack kinematics tracers up to large enough radii, where the gravitational acceleration drops substantially below the MOND acceleration constant $a_0 = 1.2\times10^{-10}$m\,s$^{-2}$, at which the MOND effects emerge. Apart from a~few exceptions, there are no objects in known orbits, similar to the gas clouds in spiral galaxies, which would enable a direct measurement of the gravitational acceleration. Jeans analysis of dynamics of stars or planetary nebulae can be used to measure the gravitational field, but its results are ambiguous since the anisotropic parameter is unknown. Gravitational lensing is not able to probe the gravitational field in the low-acceleration regime (see Sect.~1 of \citealp{milg12} for more details and a~recent review of tests of MOND in ellipticals).

\begin{figure}
\sidecaption
\resizebox{\hsize}{!}{\includegraphics{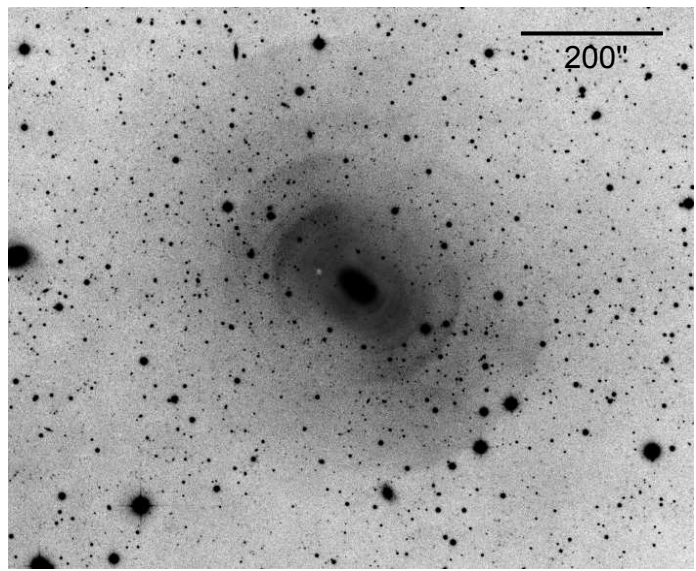}} % one column fig.
\caption{Type\,I shell elliptical galaxy NGC\,3923. Courtesy of David Malin, Australian Astronomical Observatory.}
\label{fig:N3923}
\end{figure}

Stellar shells, which are observed in many ellipticals, offer an interesting alternative to established methods to measure the gravitational field in this type of galaxies. They appear as glowing sharp-edged arcs centered on the center of the galaxy. Shells occur in about 10\% of early-type galaxies in the local Universe \citep{malin83,schweisei88,tal09,atkinson13}. In special cases, and we consider only these examples of shell galaxies hereafter, shells form an axially symmetric structure -- these are so-called Type\,I shell systems \citep{prieur90, wilkinson87}, see our Fig.~\ref{fig:N3923}. Approximately every third shell galaxy is of this type \citep{prieur90}. Several formation scenarios have been proposed, but the most accepted one today is the scenario of an almost radial minor merger \citep{quinn84}. When a~small galaxy (the secondary) reaches the center of a~bigger and more massive galaxy (the primary), the secondary is disrupted by tidal forces. Its stars begin to behave as test particles and oscillate in the potential well of the primary. When they reach their apocenters, they slow down and create kinematic density waves that are observed as the shells. However, the core of the secondary can survive the first passage. If it loses enough kinetic energy by dynamical friction, it becomes trapped in the potential well of the primary and begins to oscillate as well. During each passage through the center of the primary, the surviving part of the secondary is peeled off again and again, so that several generations of shells can be formed from one secondary. We judge that the mass ratio of colliding galaxies is about 1:10, because the total luminosity of shells usually accounts for a~few percent of the total luminosity of their host galaxy \citep{carter82,malin83-2,DC86,prieur88}. 

As we explain in Sect.~\ref{sec:shell_prop}, the radius of a~shell in a~Type\,I shell galaxy is only a~function of the shape of the galactic gravitational potential and the time elapsed since the merger. Therefore, if we know the radii of shells, we can test whether they are consistent with a~given potential. This is a~very convenient way for testing MOND, which predicts the dynamics only on the basis of the distribution of the baryonic matter. The three-dimensional density profile of the galaxy, needed to build its MOND potential, can be constrained for a shell elliptical galaxy from the morphology of its shells \citep{DC86}. Shells often extend very far from the center of their host galaxy.

The elliptical NGC\,3923 (Fig.~\ref{fig:N3923}), which is the main subject of this paper, is decorated by the biggest known shell in the Universe. Its radius is more than 100\,kpc (Table~\ref{tab:shpos}) 
and, according to the MONDian model of the galaxy NGC\,3923 from \citet{bil13}, it is exposed to a gravitational acceleration of $a_0/5$ from the host galaxy. In this work, we~theoretically predict a~new -- currently not yet observed -- shell at a distance of about 220\,kpc, which would extend to the acceleration of $a_0/10$. The modified dynamics in individual ellipticals at such a~low acceleration has been tested only in two cases -- in NGC\,720 and NGC\,1521 \citep{milg12}. 
 
Type\,I shell galaxies are interesting from the point of view of MOND for another reason as well. The modified dynamics has been tested only in systems whose constituents move in nearly circular orbits (like disk galaxies,  \citealp{thingsmond, rotcurv1, rotcurv2}), randomly distributed orbits (dwarf galaxies,  \citealp{anddwarf}, elliptical galaxies, \citealp{sanders2010}) and ellipse-like orbits (polar rings, \citealp{lughausen13}). To our knowledge, MOND has never been tested for particles in radial trajectories. Since MOND was originally inspired by disk galaxies \citep{milg83a}, it is important to test it also for strongly noncircular orbits. It is not known whether MOND should be interpreted as a~modified gravity theory or as a~modification of the law of inertia (or as a~combination of both). In the first case, the kinematic acceleration of a~test particle only depends on the vector of gravitational acceleration at the immediate position of the particle; in the latter, the kinematic acceleration can generally depend on the whole trajectory of the particle since the beginning of the Universe. In the case of modified inertia, the MOND algebraic relation $a\mu(a/a_0) = a_\mathrm{N}$ (see Sect.~\ref{sec:grid}), deduced for particles in circular orbits, is not necessarily valid for particles that oscillate along a~line. Furthermore, in many cases the stars that constitute the shells periodically travel between the Newtonian ($a\gg a_0$) and deep-MOND ($a\ll a_0$) region of their host galaxy, unlike the stars in disk galaxies, which continue to be exposed to a gravitational field of nearly constant magnitude.

The paper is organized as follows: in Sect.~\ref{sec:pospot}, we explain the equations for shell radii evolution and describe the method we use to constrain the gravitational potential of a~shell galaxy from the radial distribution of its shells. After presenting the observational data in Sect.~\ref{sec:obs_data} and our code for shell identification in Sect.~\ref{sec:aut_id}, we describe the results of the automated shell identification in Sect.~\ref{sec:res}, among which is the prediction of an as~yet undiscovered shell. The results are discussed in Sect.~\ref{sec:resdis}. In Sect.~\ref{sec:predictive_power}, we demonstrate the predictive ability of our method by reconstructing the position of a~known shell on the basis of the positions of the remaining shells. 
The functionality of the shell identification code is verified on a~Newtonian self-consistent simulation in Sect.~\ref{sec:newton}. The limitation and possible shortcomings of the method are discussed in Sect.~\ref{sec:unexpl}.
The paper is summarized in Sect.~\ref{sec:summary}.

%%%%%%%%%%%%%%%%%%%%%%%%%%%%%%%%%%%%%%%%%%%%%%%%%%%%%%%%%%%%%%%%%
\section{Relation of the shell radial distribution to the host galaxy potential}
\label{sec:pospot}
\subsection{Shell propagation} \label{sec:shell_prop}
We briefly describe the model of the shell system evolution   that we use (for more details, see Section~4 of \citealt{bil13}). A~new generation of shells is created during every passage of the secondary through the center of the primary. The maximal number of generations is, in principle, not limited, but the existing self-consistent simulations 
in the Newtonian dynamics (non-MONDian) exhibit three or four generations at most \citep{SegDup,bartoskovaselfcon,cooper11}. 
In MOND,  the merging processes tend to be more gradual than in the Newtonian dynamics with dark matter \citep{tiret08}, so that more generations can be expected to form. However, a~self-consistent MOND simulation of a~shell galaxy formation has never been published. The time difference between two subsequent passages of the secondary is also unknown and depends on the kinetic energy of the secondary and the magnitude of dynamical friction. It can only be expected that the intervals between subsequent passages will decrease because of dynamical friction. Every infall of the secondary occurs from the opposite side of the galaxy than the previous one.

We focus now on the shells from one particular generation. To derive analytic relations for the time evolution of their radii, we needed to make several simplifications. We approximated the situation by the following one-dimensional model: we assumed that the potential of the primary is distributed in a mirror symmetry around the origin of the coordinate axis. We began to measure time when the center of the secondary crossed the center of the primary.  We modeled the disruption of the secondary by releasing an infinite number of stars from the center of the primary. The velocity distribution of these stars, $f(v)$, satisfies $f(v)\neq 0$ for $v<0$, and $f(v) = 0$ for $v\geq 0$. Then the stars were let free to move in the potential well of the primary. The self-gravity of the released stars was neglected.

We denote with $t$ the time passed from the release of the stars. The first estimate of shell edge positions at the time $t$ is the radius where the stars are currently located at their apocenters. The $n$-th shell consists of the particles that reach their apocenter for the $(n+1)$-th time (or, equivalently, the $n$-th whole oscillation). The stars are located at their apocenters at the galactocentric radii $r_{\mathrm{a},n}$ if they satisfy the condition \citep{hq87}
\begin{equation}
        t = (n+1/2)P(r_{\mathrm{a},n}), 
        \label{eq:pos1}
\end{equation}
where $n$ is the so-called shell number (a non-negative integer, not to be confused with the total number of shells) and $P(r)$ denotes the period of oscillation (2\,$\times$\,the free-fall time) of a~particle with the apocenter at the galactocentric distance $r$. The function $P(r)$ is determined by the potential $\phi (r)$ of the host galaxy by the relation 
\begin{equation}
        P(r) = \sqrt{2}\int_0^r\left[\phi(r)-\phi(x)\right]^{-1/2}\mathrm{d}x.
        \label{eq:per}
\end{equation}
This approximation can differ from the precise solution of the problem by more than 10\% (see, \citealp{ebrova12}, Table~1 and 2). 

An improvement can be achieved by a~correction. According to \citet{DC86}, the shell edge moves with the velocity of stars that are located at it. Because of the conservation of energy of every individual particle, there is a~relation between the position of the $n$-th shell $r_{\mathrm{s},n}$, the position of the apocenters $r_{\mathrm{a},n}$ of stars making up the shell, and the shell phase velocity $v_{\mathrm{s},n}$: 
\begin{equation}
        \phi (r_{\mathrm{s},n})+\frac{1}{2}v_{\mathrm{s},n}^2 = \phi(r_{\mathrm{a},n}). 
\end{equation}
The position of the shell is then 
\begin{equation}
        r_{\mathrm{s},n} = (-1)^{n+1}\phi^{-1}\left[ \phi(r_{\mathrm{a},n}) -\frac{1}{2}v_{\mathrm{s},n}^2 \right], 
        \label{eq:pos2}
\end{equation}
where $\phi^{-1}$ denotes the inversion function of $\phi$. The first factor on the right-hand side expresses the fact that shells of odd number tend to lie on the opposite side of the galaxy from those with even numbers. By accepting this sign convention we assumed that the secondary flew in from the positive side of the collision axis (the center of the primary lies at the origin of the axis). The analytic expression for $v_{\mathrm{s},n}$ is in general not known. We approximated it by the phase velocity of the respective apocentric radius
\begin{equation}
        v_{\mathrm{s},n} \approx v_{\mathrm{a},n} = \frac{1}{\left(n+1/2\right)\frac{\mathrm{d}P(r)}{\mathrm{d}r}\left|_{r = r_{\mathrm{a},n}}\right.}.
        \label{eq:vsva}
\end{equation}
These equations excellently describe the shell propagation in test-particle simulations (like those of \citealt{hq87} or \citealt{ebrova12}). 

If the shell systems were formed in several generations and the age of the $N$-th generation is $t_N$, the set of shell radii can be expressed as
\begin{eqnarray}
  \left\{ o_I (-1)^{N+1}r_{\mathrm{s},n}(t_N) ;\ N=1, 2,\ldots, N_\mathrm{max}; \right. \nonumber \\ 
  \left. n = n_{\mathrm{min}_N}, n_{\mathrm{min}_N}+1, \ldots ,n_{\mathrm{max}_N} \right\},
\label{eq:rall}
\end{eqnarray}
where the generation number $N$ extends over all generations present in the shell system and $n$ over all shell numbers from the $N$-th generation. The sign of the first generation $o_I$ can be $+1$ or $-1$, depending on whether the secondary flew in from the positive or negative part of the collision axis. For an observed Type\,I shell galaxy, the collision axis is also the symmetry axis of the shell system and the terms positive and negative part of the axis are a~matter of convention.

The shell edges form approximately spherical caps centered on the nucleus of the host galaxy. The ellipticity of shell edges in the sample of observed and simulated shell galaxies of \citet{DC86} ranges from 0 for E0 galaxies to 0.3 for E7 galaxies. The projected radius of the shell therefore weakly depends on the angle of view of the observer. But if the observer is situated near the axis of the shell system, the shells would not appear as sharp-edged features.

Stellar tails are often present in shell galaxies \citep{tal09,duc11,ramos11}. 
In simulations of radial minor mergers, the formation of each generation of shells is accompanied by a formation of one tail. 
The tail is located on the opposite side of the galaxy from that from which the secondary originally came.
The stars in the tail escape from the system or perform the first oscillation from pericenter to apocenter and back. The surface brightness of the tail decreases with time as more and more stars complete the first oscillation. No tail has been reported in NGC\,3923 so far. 

The zeroth shell ($n=0$) consists of stars in their first apocenter after they were released from the secondary. It is part of the stellar tail. The zeroth shell is not considered as a~shell by some authors because it is not a~phase-space wrap like the other shells \citep{quinn84}. Its morphology depends on many factors. In simulations, it can look like an ordinary shell (e.g., in the simulation of \citealp{cooper11}), it can be diffuse, or even not observable. But its surface brightness is comparable with, or lower than, that of the stellar tail.

To summarize: Each time the secondary survives the passage through the center of the primary, a~new generation of shells is produced. Several shells are produced in a~generation. Their number depends on the spectrum of kinetic energies of the stars at the moment of their release, on the profile of the primary potential, on the time since the passage (the age of the generation), and our ability to detect them. If the energy spectrum of the stars released in a~generation is single-peaked, then a~smallest $r_\mathrm{min}$ and largest $r_\mathrm{max}$ radius exists between which the shells are observable. Shells increase their radius with time. Shells of lower number $n$ propagate faster than shells with higher $n$ at the same radius. New shells form at $r_\mathrm{min}$ and spread until they reach $r_\mathrm{max}$ , where their surface brightness decreases below the detection limit. Another mechanism of shell-fading exists: as the age of the shell system increases, more shells are formed, which means that only a~small fraction of stars forms shells of high shell number, which in turn
results in the low surface brightness of these shells. One stellar tail is associated with each generation of shells.

%%%%%%%%%%%%%%%%%%%%%%%%%%%%%%%%%%%%%%%%%%%%%%%%%%%%%%%%%%%%%%%%%
\subsection{Shell identification} \label{sec:id}
There were attempts to use Eqs.~(\ref{eq:pos1}) and (\ref{eq:per}) (i.e., the shell position was approximated by the radius where the stars are just in their apocenter) to constrain the potential of Type\,I shell galaxies in the 1980s \citep{quinn84, hq87}. The common assumption at that time was that the whole shell system is formed in one generation. It was found later that single-generation simulations were unable to reproduce the high radial range 
(which is the ratio of the outermost and the innermost shell) observed in many shell galaxies \citep{DC86}. 
This is also the case of NGC\,3923, which is the galaxy with the highest known shell radial range (around 60). Some of the stars had to lose orbital energy somehow to form shells close to the galactic nucleus. The analytical estimations of \citet{dupcomb87}, taking into account the mass loss of the secondary and dynamical friction acting on it, revealed the necessity of formation in multiple generations. Finally, this way of shell formation was observed in numerous self-consistent simulations \citep{salmon90,SegDup,bartoskovaselfcon,cooper11}.

Considering an image of a~shell galaxy, the generation of a~particular shell and its shell number (the identification of the shell) are not known a priori. Some shells might also be missed by observations. These results caused the decline of interest in probing the galactic potential using shell radial distribution at the end of the 1980s
because all methods used at that time were based on the single-generation origin of the whole shell system. 

The first solution of these problems appeared in \citet{bil13}. Our method allows testing the compatibility of an observed radial distribution of shells with a~given gravitational potential. The method is  unable to reconstruct the potential unambiguously. However, it can be applied to a~grid of potentials to select a~preferred region in the grid. The method works as follows: first, we add a~sign to the measured shell radii. If, for example, the axis of the shell system lies in the north-south direction, we add a plus sign to the radii of all shells on the northern side of the galaxy and a minus sign to the radii of all shells on the southern side. To calculate the time evolution of shell radii, one uses Eqs.~(\ref{eq:pos1})\,--\,(\ref{eq:pos2}) (or a better model). The next step is to assign a~shell number and a~generation number to each of the observed shells, that is, to perform the shell identification. However, the identification must satisfy several criteria to be plausible. The criteria are that
\begin{itemize}
        \item the observed shell radii must be well reproducible by the model, meaning that for appropriate ages $t_N$, the observed shell radii are close to those given by Eq.~(\ref{eq:rall}). The sign of the first generation $o_I$ will be $+1$ if we assign an even shell number and the first generation to an observed shell with positive radius (i.e., the shell at the northern side of the galaxy, in our previous example). In the opposite case, $o_I$ will be $-1$.
        \item the identification should not require many missing shells. This means that for every generation, the shell numbers should form a~continuous series of integers (e.g., 3, 4, 5, 6). If a~number is missing in this series, the corresponding shell must be escaping observations. If shells number $m$ and $n$ are observed, but shells with numbers between $m$ and $n$ are not, we call these unobserved shells the missing shells. If a~sufficient number of shells is allowed to be missing, it is easy to fulfill the remaining criteria for almost any arbitrary combination of shell radii and a~potential. 
        \item not many generations are allowed for the same reason.
        \item the $N$-th generation must be older than the $(N+1)$-th generation for every $N$.
        \item the differences between ages of subsequent generations are becoming shorter because the amplitude of oscillations of the secondary is damped by dynamical friction. 
\end{itemize}
If such an identification exists, then the observed shell radii can be considered compatible with the given potential. The choice of the identification criteria is nontrivial. Our choice and its motivation is described in Sect. \ref{sec:aut_id}.

In \citet{bil13}, we applied this test to the shells of NGC\,3923 and its MONDian potential, assuming MOND as a~modified gravity theory. We found an acceptable identification for the 25 outermost shells of the galaxy (out of 27). They were explained by three generations of shells. The deviations of the modeled shell radii from the observation were lower than 5.4\%. At that time, we were not able to say whether the identification of at least some of the shells was unequivocal. This was the reason for creating the code for automatic shell identification.

%%%%%%%%%%%%%%%%%%%%%%%%%%%%%%%%%%%%%%%%%%%%%%%%%%%%%%%%%%%%%%%%%
%%%%%%%%%%%%%%%%%%%%%%%%%%%%%%%%%%%%%%%%%%%%%%%%%%%%%%%%%%%%%%%%%
\section{Observational data} \label{sec:obs_data}
Our MOND models of the gravitational potential of NGC\,3923 are based on the surface brightness profile of the galaxy derived in \citet{bil13}. In that paper, we used the data from the Spitzer telescope in the 3.6\,$\mu$m band. The profile was constructed by fitting the image of the galaxy by a~sum of two S\'ersic profiles. In that paper, we derived an effective radius of 235 (7.58)$\,\arcsec$, an effective surface brightness of 20.5 (15.9) mag\,arcsec$^{-2}$, and a S{\' e}rsic index of 5.28 (1.53) for the first (second) component. The components shared the same ellipticity of 0.316 (defined as $1-a/b $). 

In the present paper, we assumed that the galaxy is a~prolate ellipsoid. This was motivated by two facts: a) minor axis rotation was detected by \citet{minrot} 
and b) the simulations of \citet{DC86} showed that a~shell systems with the morphology of NGC\,3923 can form only in prolate potentials. The galaxy was assumed to be viewed perpendicularly to its major axis. We treat MOND as a~modification of gravity, where the acceleration of a~star depends on the gravitational field at its immediate position.

In \citet{bil13} we derived the mass-to-light ratio in the Spitzer 3.6\,$\mu$m band as 0.78 and we adopted the distance of the galaxy from Earth of 23\,Mpc. All the direct measurements of the galactic distance in the NASA/IPAC Extragalactic Database range from 19.9 to 24\,Mpc. The results of \citet{sauronXIX} for the mass-to-light ratio in the 3.6\,$\mu$m band varied from 0.3 to 1.21 in their sample of 48 E and S0 galaxies. 

The shell radii adopted from \citet{bil13} are given in Table~\ref{tab:shpos}. The symmetry axis of the shell system of NGC\,3923 lies in the NE-SW direction. We added the  plus sign to shells located on the north-eastern side of the galaxy and the minus sign to those on the southwestern side. The label is an arbitrary name of a~shell. We found it more convenient in the context of the multigeneration scenario of shell formation than the traditionally used observational number, which was motivated by the hypothesis of formation of shells in a~single generation.
Just as in \citet{bil13}, we assume that the observed shell with label $i$ is shell $h$ that encircles the galaxy. The influence of this assumption on our results is discussed in Sect.~\ref{sec:2sh}.

\begin{table}
\caption{Shells of NGC\,3923}\label{tab:shpos}
\centering
\begin{tabular}{crrr}
\hline\hline
Label & $r$ $[\arcsec]$ & $R$ $[$kpc$]$ & $a$ $[a_0]$ \\
\hline
$a $ &  $+1170$ &       $130.7$ &       $0.2$ \\
$b $ &  $-840$ &        $93.8$ &        $0.3$ \\
$c $ &  $+630$ &        $70.4$ &        $0.4$ \\
$d $ &  $-520$ &        $58.1$ &        $0.4$ \\
$e $ &  $+365$ &        $40.8$ &        $0.6$ \\
$f $ &  $-280$ &        $31.3$ &        $0.8$ \\
$g $ &  $+203$ &        $22.7$ &        $1.1$ \\
$h $ &  $-148.5$ &      $16.6$ &        $1.5$ \\
$i $ &  $+147.3$ &      $16.5$ &        $1.5$ \\
$j $ &  $+128.1$ &      $14.3$ &        $1.7$ \\
$k $ &  $-103.6$ &      $11.6$ &        $2.2$ \\
$l $ &  $+99.9$ &       $11.2$ &        $2.2$ \\
$m $ &  $-79.6$ &       $8.9$ & $2.9$ \\
$n $ &  $+72.8$ &       $8.1$ & $3.1$ \\
$o $ &  $-67.0$ &       $7.5$ & $3.4$ \\
$p $ &  $+64.1$ &       $7.2$ & $3.6$ \\
$q $ &  $+60.4$ &       $6.7$ & $3.9$ \\
$r $ &  $-55.5$ &       $6.2$ & $4.2$ \\
$s $ &  $+51.2$ &       $5.7$ & $4.6$ \\
$t $ &  $-44.0$ &       $4.9$ & $5.5$ \\
$u $ &  $+41.5$ &       $4.6$ & $5.9$ \\
$v $ &  $-37.7$ &       $4.2$ & $6.6$ \\
$w $ &  $+34.3$ &       $3.8$ & $7.3$ \\
$x $ &  $+29.3$ &       $3.3$ & $8.8$ \\
$y $ &  $-28.7$ &       $3.2$ & $9.0$ \\
$A $ &  $+19.4$ &       $2.2$ & $14$ \\
$B $ &  $-18.0$ &       $2.0$ & $15$ \\
\hline
\end{tabular}
\label{tab:sh}
\tablefoot{
$r$\,--\,the observed distance of the shell from the center of NGC\,3923, data compilation taken from \citet{bil13}. The plus sign means that the shell is situated on the northeastern side of the galaxy, the minus sign that it is on the southwestern
side; $R$\,--\,the radius of the shell in kiloparsecs assuming a galaxy distance of 23\,Mpc; $a$\,--\,the gravitational acceleration at the radius of the shell for the potential derived in \citet{bil13}.}
\end{table}

%%%%%%%%%%%%%%%%%%%%%%%%%%%%%%%%%%%%%%%%%%%%%%%%%%%%%%%%%%%%%%%%%
%%%%%%%%%%%%%%%%%%%%%%%%%%%%%%%%%%%%%%%%%%%%%%%%%%%%%%%%%%%%%%%%%
\section{Automated shell identification} \label{sec:aut_id}
We developed a~code that automatically performs the procedure described in Sect.~\ref{sec:id}. The code takes the set of the observed shell radii and the tested potential as the input and returns all the identifications of the shells that meet our criteria. If no identification can be found, the potential is considered incompatible with the observed shell distribution. The identification criteria are defined by choosing these parameters: 
\begin{enumerate}
        \item The largest allowed number of generations $N_\mathrm{max}$.
        \item The largest allowed number of missing shells in every generation $n_\mathrm{miss}$. 
        \item The largest allowed deviation $\Delta$ of the observed shell radius from the corresponding modeled shell (the tolerance parameter). If the shell labeled $\lambda$ is identified as shell the $n$ from the $N$-th generation and the age of the $N$-generation is $t_N$, we define its deviation from the model as the relative difference  $\delta_\lambda = \left|o_I(-1)^{N+1}r_{\mathrm{s},n}(t_N)/r_\lambda-1\right|$, where $r_{\mathrm{s},n}(t_N)$ is calculated from Eqs.~(\ref{eq:pos1})\,--\,(\ref{eq:pos2}). An identification is not considered satisfactory unless there are $t_I, t_{II}, t_{III}, \ldots, t_{N_\mathrm{max}}$ such that $\delta_\lambda\leq \Delta$, for all the observed shells.
        \item The oldest allowed age of the first generation (i.e., the time from the first encounter of the centers of the primary and the secondary). 
        \item The largest allowed shell number.
\end{enumerate}
Unless stated otherwise, we looked for the identification only the 25 outermost shells of 27 (excluding $A$ and $B$), considered shells $h$ and $i$ as one shell with a radius of $h$ (as explained in \citealp{bil13}) and used these  parameters:
\begin{itemize}
        \item $N_\mathrm{max} = 3$. If we assume that the real shells evolve so that their radii do not deviate from the model given by Eqs.~(\ref{eq:pos1})\,--\,(\ref{eq:pos2}) by more than 10\%, at least three generations are necessary to explain the observed shell distribution of NGC\,3923 in the potential from \citet{bil13}.
        \item For the same reason, we set $n_\mathrm{miss} = 2$ for every generation. 
        \item We set $\Delta$=7\%. The motivation for this value is explained in Sect.~\ref{sec:grid}. 
        \item We restricted the age of the shell system by the value of $cP(r_a)$, where $P(r_a)$ denotes the period of oscillation, Eq.~(\ref{eq:per}), at the radius of the outermost shell (label $a$ in Table~\ref{tab:shpos}) in the given potential. The constant $c$ was set to 3. This choice is also explained in Sect.~\ref{sec:grid}.
        \item The largest allowed shell number was chosen to be 29. This is much more than the largest shell number produced in any published test-particle simulation.
\end{itemize}
An observed shell can be identified as two or more overlapping shells if the position of the modeled shells is closer than the tolerance $\Delta$ from the radius of the observed shell. The direction of the initial infall of the secondary (i.e. the sign of the first generation, $o_\mathrm{I}$) is determined by the chosen identification (see Sect.~\ref{sec:id}).

%%%%%%%%%%%%%%%%%%%%%%%%%%%%%%%%%%%%%%%%%%%%%%%%%%%%%%%%%%%%%%%%% 
%%%%%%%%%%%%%%%%%%%%%%%%%%%%%%%%%%%%%%%%%%%%%%%%%%%%%%%%%%%%%%%%%
\section{Results}
\label{sec:res}
\subsection{Basic grid of gravitational potentials} 
\label{sec:grid}
The classical free parameters in MOND are the mass-to-light ratio and the distance of the galaxy from Earth. In our basic grid of potentials, every node corresponds to a~pair of values of the mass-to-light ratio in the 3.6\,$\mu$m band, $M/L_{3.6}$, and the distance of NGC\,3923, $d$. The variable $M/L_{3.6}$ takes values between 0.3 to 1.21 in our grid (lowest and highest values found in a~large sample of early-type galaxies, see Sect. \ref{sec:obs_data}) in 50 equidistant steps. The galactic distance $d$ is sampled in the same number of steps from 19 to 25\,Mpc (broader than the observational limits). For every node of the grid, the density profile and the shell radii (in kiloparsecs) are adjusted so that the galaxy photometric profile and the shell radii (in arcseconds), as observed from Earth, do not change. By assuming that the ellipsoid is viewed from its symmetry plane, Abel deprojection can be used and the volume density has the same axis ratio as the surface density. The Newtonian acceleration $a_\mathrm{N}$ along the ellipsoid major axis (which coincides with the symmetry axis of the shells) was calculated by our own code. The correctness of the code was successfully verified on Ferrer's ellipsoids with known analytical potentials \citep{binney-tremaine08}. 
The Newtonian gravitational acceleration $a_\mathrm{N}$ was converted to the MONDian one, $a_\mathrm{M}$, using the algebraic relation \citep{famaey12}
\begin{equation}
        a_\mathrm{M}\,\mu\left(\frac{a_\mathrm{M}}{a_0}\right) = a_\mathrm{N},
        \label{eq:anam}
\end{equation}
where $\mu$ is the ``simple'' interpolation function
\begin{equation}
        \mu(x) = \frac{x}{1+x},
        \label{eq:musimple}
\end{equation}
and $a_0=1.2\times 10^{-10}$m\,s$^{-2}$ is the MOND acceleration constant. We call this set of potentials the basic grid.

\begin{figure}
\sidecaption
\resizebox{\hsize}{!}{\includegraphics{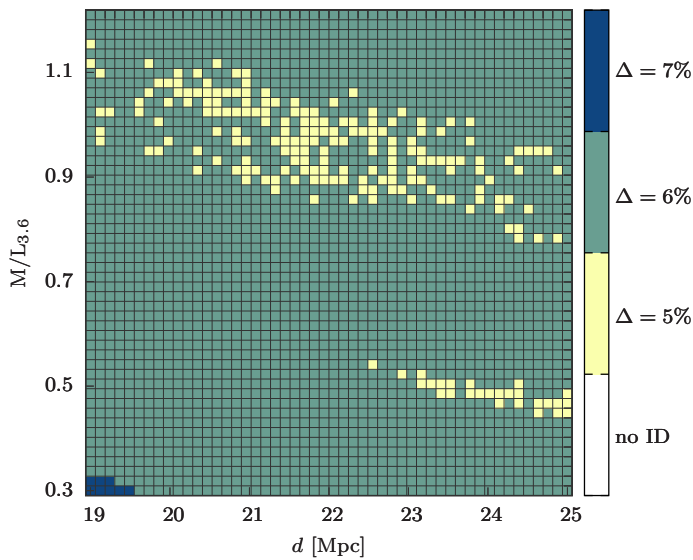}} % one column fig.
%\resizebox{\hsize}{!}{\includegraphics{fig_profile.eps}} include .eps into the final version, but .pdf to the draft versions (eps->pdf by epstopdf --autorotate=All fig_*.eps
\caption{Map of the gravitational potentials from the basic grid that are compatible with the observed shell radial distribution of NGC\,3923. Different colors correspond to different choices of the tolerance parameter $\Delta$. A~potential is considered compatible with the shell distribution if the positions of all the observed shell radii differ by less than $\Delta$\% from those modeled in the respective potential.}
\label{fig:mapa_simple}
\end{figure}

Now explain our choice of the tolerance parameter $\Delta$. It is not possible to find a~shell identification for $\Delta=4$\% for any potential from the basic grid. The situation changes for $\Delta=$\,5\%, when a~few successfully identified nodes appear (Fig.~\ref{fig:mapa_simple}). 
These potentials are only consistent with the observed shell radii if the four outermost shells of NGC\,3923 have shell numbers 2, 3, 4, and 5 and they originate in the first generation. When $\Delta$ is increased, more and more identifications are found, but the identification of the four outermost shells remains unchanged. The break comes at $\Delta = 10$\% (we explored only the integer values of $\Delta$). Then another identification of the four shells appears, which gives them shell numbers 3, 4, 5, and
6. However, if one model reproduces the observations twice as well as the other model, one should prefer the better one. 
Moreover, in the self-consistent Newtonian simulation described in Sect.~\ref{sec:newton}, our code was able to find the correct identification for $\Delta = 4$\%. For these reasons we considered a 10\% tolerance too high.

For every potential from the basic grid that was compatible with the observed shell radial distribution, at least one of the found identifications assigns the shell numbers 2, 3, 4, and 5 and the first generation to the four outermost shells of NGC\,3923 (shells $a$, $b$, $c$, $d$ in the Table~\ref{tab:shpos}). For the three outermost shells no satisfactory other identification exists. This means that there must be a shell number~1. We estimate its position in this work. To give reliable error bars for this shell, we chose a~high value of $\Delta$. Since the potentials at the edges of our basic grid are very improbable, we chose the lowest possible $\Delta$ so that an identification can be found for every potential from the basic grid. 
That is $\Delta = 7$\%. 

In the criterion for the oldest allowed age of the first generation, $cP(r_a)$, we set $c=3$ because no new identifications were found for $c= 10$ in the sample of 50 randomly chosen potentials from the basic grid. In fact, the required values of $c$ was between 2.5 and 2.6 for all of the randomly chosen potentials. For example, for the potential of \citet{bil13} $P(r_a) = $1107\,Myr. For old ages of the shell system, a~densely spaced shell distribution is predicted by our model, which would require to postulate many
missing shells.

In Fig.~\ref{fig:mapa_simple}, we can see the map of potentials in the basic grid for which an identification was found for the given values of $\Delta$.

\begin{figure*}
\sidecaption
\resizebox{0.9\hsize}{!}{\includegraphics{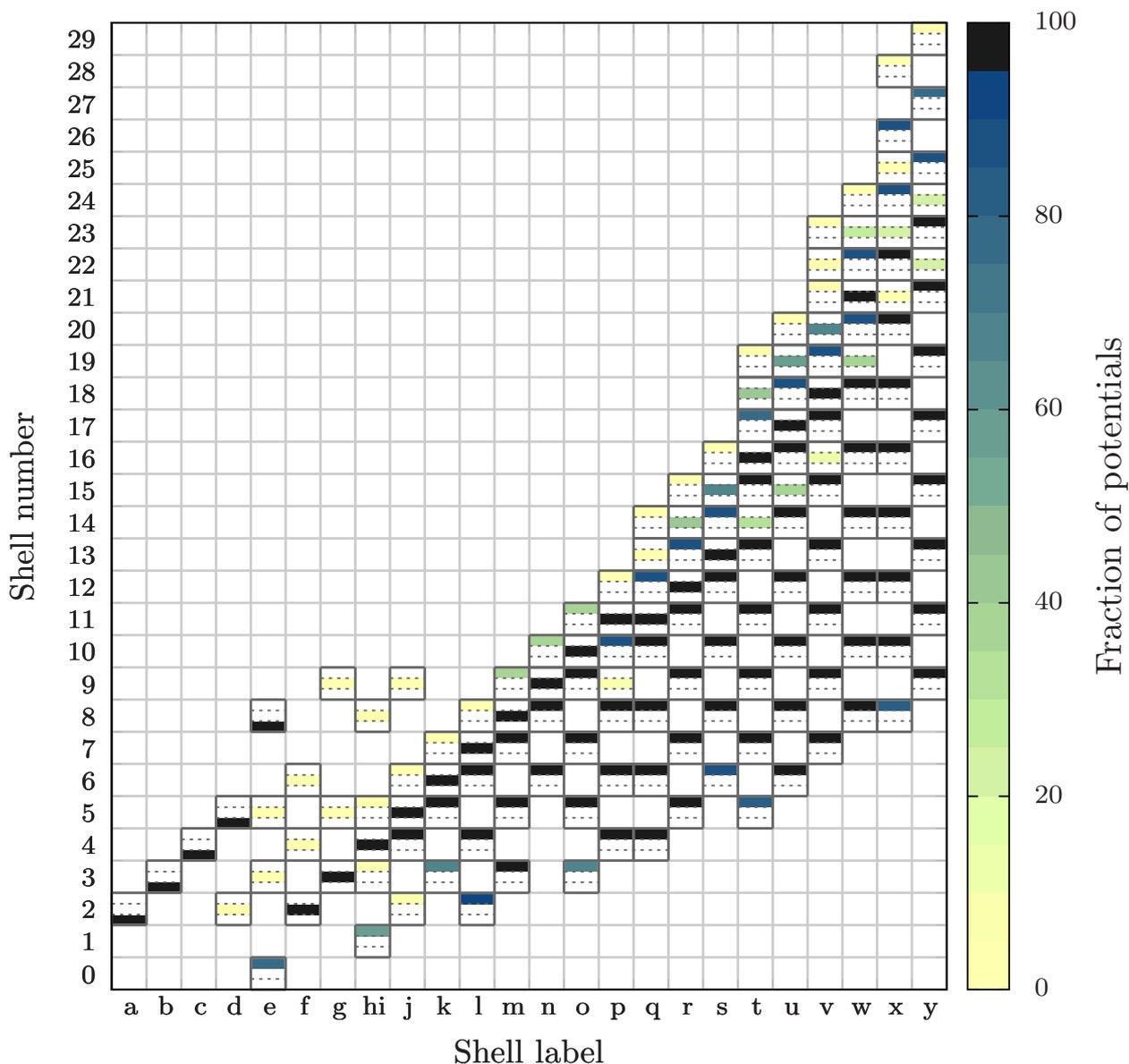}} % one column fig.
%\resizebox{\hsize}{!}{\includegraphics{fig_shcomp.eps}} include .eps into the final version, but .pdf to the draft versions (eps->pdf by epstopdf --autorotate=All fig_shcomp.eps
\caption{All satisfactory identifications found in our basic grid of gravitational potentials that meet the identification criteria defined in Sect.~\ref{sec:aut_id}. An identification of a~shell is the assignment of a~shell number and a generation number to it. The subcells of every cell denote the identified generation of the shell -- bottom part: the first generation; middle part: the second generation; upper part: the third generation. The color of the subcell encodes the number of the potentials from the basic grid (Sect.~\ref{sec:grid}) for which the shell was successfully identified in the given way, divided by the total number of potential in the basic grid. }
\label{fig:idtab_basgrid}
\end{figure*}

All shell identifications for the potentials of the basic grid are shown in Fig.~\ref{fig:idtab_basgrid}.
The horizontal axis shows the label of the shell (as in Table~\ref{tab:shpos}), the vertical axis the identified shell number. The subcells represent the identified generation: first\,--\,bottom part, second\,--\,middle part, or third\,--\,top part of the cell. The color stands for the fraction of the potentials for which the shell is assigned the identification divided by the total number of potentials in the grid. In some cases, several identifications of a shell were compatible with one potential, therefore the sum of fractions in one column can exceed 100\%. 

The identifications of the inner shells (those on the right) are highly ambiguous. This is because the observed shells are relatively close to each other compared to the chosen value of $\Delta$. Moreover, the tested potentials produce densely spaced shells near the center because the oscillation period is much shorter near the center than at the large radii. 

The situation is much better for the four outermost shells. For every potential in the basic grid, they can be identified as shells number 2, 3, 4, and 5 from the first generation. This is the main result of this work. It implies an observable prediction: shell number 1 must exist. In the following, we try to determine its position and discuss whether this result changes when some of our assumptions are modified. 

The next step was to estimate the position of the new shell. For every node of the basic grid, we found the time $t_I$ at which the four outermost shells that belonged to the first generation are fitted best by the model (i.e., the time at which the maximum of the relative difference over the four outermost shells is minimized). For this value of the time we calculated the position of shell number 1. By applying this procedure for every node of the basic grid, we obtained a~set of predicted radii of the new shell. The statistics of this set can be found in the first row of Table~\ref{tab:possh1}.

\begin{table}
        \centering
\caption{Statistics of the predicted radius of the new shell.}
                \begin{tabular}{lcccc}\hline\hline
        & $     r_{1,\mathrm{median}}$          & RMS           & $r_{1,\mathrm{min}}$    & $r_{1,\mathrm{max}}$\\\hline
basic grid        & 1904        & 40    & 1850 &        2025\\
``standard'' $\mu$ (Eq.~\ref{eq:musta})  & 2000 & 48    & 1918  & 2092\\
$a_0'=0.7a_0$            & 1880 & 35    & 1841  & 2001\\
$a_0'=1.3a_0$    & 1931 & 43    & 1861  & 2040\\
expanded density profile        & 1970 & 20 &   1930 & 2024\\
point mass        & 1774        & 17    &       1762    & 1814\\\hline
                \end{tabular}
                \tablefoot{Statistics of the position of the predicted shell -- median, standard deviation (RMS), minimum, and maximum. All quantities are in arcseconds.}
                \label{tab:possh1}
\end{table}

%%%%%%%%%%%%%%%%%%%%%%%%%%%%%%%%%%%%%%%%%%%%%%%%%%%%%%%%%%%%%%%%%
\subsection{Varying the interpolation function} \label{sec:interpol}
The MOND potential, of course, depends on the choice of the interpolation function. Another favorite interpolation function is the ``standard'' function
\begin{equation}
        \mu(x) = \frac{x}{\sqrt{1+x^2}}.
        \label{eq:musta}
\end{equation}
We repeated the procedure from the previous section, but replaced the interpolation function by this one. The identification criteria remained the same. The results did not change much: three generations were needed to explain the position of the 25 outermost shells of NGC\,3923 and it was possible for all the potentials compatible with the observed shell position to assign shell numbers 2, 3, 4, and 5 to the four outermost shells and sort them into the first generation. The position of the new shell was be 2000$\arcsec$ (see the second row of Table~\ref{tab:possh1}).

\subsection{Another value of $a_0$}
The value of the acceleration constant $a_0$ is not known precisely. Various authors derived $a_0$ with an uncertainty of a~few tens of percent (e.g, \citealp{thingsmond, mcgaugh11}). The value of $a_0$, that works best in the approximative algebraic relation~(\ref{eq:anam}), also depends on the distribution of mass \citep{famaey12}. We therefore tested how our results change with the variation of $a_0$ and found that they do not vary much. The basic grid of potentials was calculated again but now with a~different value of $a_0$. We checked 70\% and 130\% of the canonical value of $a_0$ (Sect.~\ref{sec:grid}). We again  used the simple interpolation function and the same identification criteria as before.
The identification of the four outermost shells did not change in either case. The positions of the new shell are given in the third and fourth rows of Table~\ref{tab:possh1}.

%%%%%%%%%%%%%%%%%%%%%%%%%%%%%%%%%%%%%%%%%%%%%%%%%%%%%%%%%%%%%%%%%
\subsection{Another luminosity profile}
\label{sec:lum}
Although the fit of the luminosity profile of NGC\,3923 (\citealt{bil13}, their Fig.~2) 
goes through the measured data, we here discuss the influence of the change of the mass distribution on the identification of the outer shells. The actual mass distribution of the galaxy may be different from what we assumed so far, for example, because the three-dimensional shape or orientation of the galaxy in space is different, or its mass-to-light ratio varies substantially within its volume.
On the other hand, the outer shells may be so distant that their positions are largely determined by the asymptotic behavior of the potential. To check this out, we replaced the fitted luminosity distribution by a~completely different luminosity distribution. We made two new potential grids. The grids were again similar to the basic grid, only the luminosity profile of the galaxy was changed. It was replaced, in the first case, by the same luminosity profile, but its space scale was expanded by a factor of two and, in the other case, by a~point mass. The observed total luminosity was conserved.
We searched for the identification of only four outermost shells and only one generation was allowed (the inner shells would be identified incorrectly with these luminosity distributions anyway). 
The other parameters of the code were left as before. 

The results were the same. All identifications found by the code assigned shell numbers 2, 3, 4, and 5 and the first generation to the four outermost shells. 
The predicted distance of shell 1 is listed in the fourth and fifth rows of Table~\ref{tab:possh1}.

%%%%%%%%%%%%%%%%%%%%%%%%%%%%%%%%%%%%%%%%%%%%%%%%%%%%%%%%%%%%%%%%%
\subsection{Two shells instead of one} \label{sec:2sh}
Until now, we excluded shell $i$ from identifications because we interpret it as shell $h$ that encircles the galaxy \citep{bil13}. Now, we allowed the $h$ and $i$ shells to be two independent shells, and we ran our identification code on the basic grid of potentials. The result was that it is possible to find an identification for our identification criteria (Sect.~\ref{sec:aut_id}), but for the price that at least two shells are missing in each generation. When shells $h$ and $i$ are treated as one shell, most of the identifications need only one or zero missing shells in the second and third generation. Nevertheless, the identification of the outer four shells remains unchanged if shells $h$ and $i$ are treated as two individual shells. 

\section{Discussion}
\label{sec:disc}
\subsection{Discussion of the results}
\label{sec:resdis}
In Sects.~\ref{sec:grid}--\ref{sec:2sh} we varied some of the free parameters of the problem of shell identification in NGC\,3923. Here we summarize and discuss the results.

In all the potentials we have explored and that are compatible with the observed shell radial distribution, it is possible to identify the four outermost shells ($a$, $b$, $c$ and $d$) as shells number 2, 3, 4, and 5 from the first generation. Moreover, the identification of the three outermost shells cannot be different. This leads us to expect the existence of shell number 1. However, it may be very faint. As we said in the Introduction (Sect.~\ref{sec:intro}), the shells disappear when they become too old and big because their surface brightness decreases under the detection limit. The surface brightness depends, among others, on the energy spectrum of the stars released from the secondary during its passage through the center of the primary. At the moment, we can only make the following rough estimate: if we assume that the new predicted shell  constitutes the same number of stars as shell $a$, and the radius of the former is approximately twice as large as the latter, its surface brightness must be four times lower. The
surface brightness of shell $a$ is about 26.5\,mag\,arcsec$^{-2}$ in the B band, which means that the new shell should have about 28\,mag\,arcsec$^{-2}$. This value is in the reach of existing instruments.

All the identifications have the common feature that the observed shell distribution of NGC\,3923 can be explained in MOND only if the shell system was created in at least three generations (if a~shell distribution is compatible with the formation in $N$ generations, then it is evidently compatible with any number of generations higher than $N$). The three generations are sufficient to account for 25 out of 27 shells. The innermost two shells probably originate from the fourth generation, but the secondary remainder could have decayed before it reached the center of the primary for the fourth time. 

In simulations, each passage of the secondary through the primary center is accompanied by the formation of one stellar tail. Therefore three tails could be present in NGC\,3923, but their surface brightness can be low. Since all the identifications imply that the secondary had to impact NGC\,3923 along the southwestern major semi-axis at their first collision, two of the tails should point to the northeast and one to the southwest.

In Table~\ref{tab:possh1} we can see the influence of the individual free parameters on the position of the predicted shell. The choice of the interpolation function has a similar influence on its radius as the variations of $M/L_{3.6}$ and $d$. Moreover, $M/L_{3.6}$ and $d$ were intentionally left to vary beyond the observational constraints. On the other hand, the choice of $a_0$ affects the radius of the expected shell much less than the choice of the interpolation function. 

Substituting the real galaxy by a point mass seems to be a severe modification that does not produce a meaningful 
 prediction of the shell position. However, we can learn from this example that substituting a~potential for a~more concentrated one leads to an inward shift of the position of the predicted shell (Table~\ref{tab:possh1}). In contrast, the twice-expanded mass distribution is not as meaningless. If NGC\,3923 is not viewed perpendicularly to its major axis, the galaxy is more elongated than we have assumed. If it is the case, the predicted shell can be expected at a larger radius. 

The most probable radius of the predicted shell is the radius derived for the most probable combination of the mass profile, mass-to-light ratio, and galaxy distance from Earth, which are those derived in \citet{bil13}. This radius is +1874$\arcsec$ for the ``simple'' interpolation function, and +1940$\arcsec$ for the ``standard'' interpolation function. 

The ages of the generations are quite sensitive to the choice of the free parameters. For example, for the basic grid, the age of the first generation ranges from 2 to 4\,Gyr, the age of the second from 400 to 2200\,Myr, and that of the third generation even from 100 to 1000\,Myr. For the potential and the shell identification used in \citep{bil13}, the ages of the first, second, and third generation are derived as 2688, 631 and 364\,Myr, respectively.

We did not designate shell $e$ as shell number 8 from the first generation (as suggested by Fig.~\ref{fig:idtab_basgrid}) and did not search for shell number 1 of the second generation. This may be related. We hesitated to consider shell $e$ as shell 8 of the first generation because it is clearly visible in the images of the galaxy, but shells number 6 and 7 are not. This situation never occurred in our simulations of shell galaxy formation (see, e.g., \citealp{ebrova12}), but we have experience only with exactly radial mergers and spherical primaries. Maybe this is not a~serious problem, because the surface brightness of individual shells in photographs of NGC\,3923 shows no radial trend and it is rather random. 

The other disturbing fact is that shell $f$, which is almost always classified as shell number 2 of the second generation, is very bright, but a~shell with a radius compatible with that
of shell number 1 of the second generation has never been reported. This shell should lie at a radius of about +450$\arcsec$. Possibly, shell $e$ is shell 1 of the second generation, which was shifted inward by some effect. The shift can be easily explained if the material that formed the shell was released before the secondary reached the primary's center. We searched for a~photograph of the galaxy that would map the region of the expected occurrence of shell 1 of the second generation, but we were not able find one. We did not even find an~image that clearly showed shell $d$. The region of the galactocentric distances 400 and 600$\arcsec$ has probably never been properly explored. This region lies between the area covered by the modern narrow field-of-view CCD observations that show the central part of the galaxy and of the older wide-field images taken on photographic plates, which were processed to show the two faint outermost shells at the expense of saturating the center of the galaxy.

\begin{figure}[b!]
\sidecaption
\resizebox{\hsize}{!}{\includegraphics{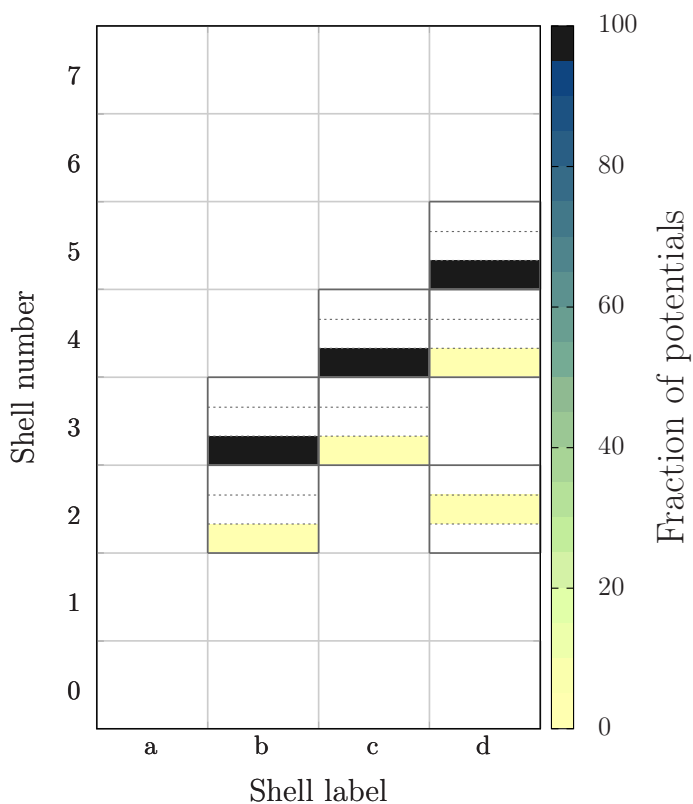}} % one column fig.
%\resizebox{\hsize}{!}{\includegraphics{fig_shcomp.eps}} include .eps into the final version, but .pdf to the draft versions (eps->pdf by epstopdf --autorotate=All fig_shcomp.eps
\caption{Test of the predictive power of our method. Shell $a$ was omitted from the set of shell radii we used elsewhere in this paper and the procedure from the Sect.~\ref{sec:grid} was repeated. The meaning of this figure is analogous to that of Fig.~\ref{fig:idtab_basgrid}. For all the potentials from the basic grid, shells $b$, $c$ and $d$ can be identified as shells number 3, 4 and 5 from the first generation. If we did not know about the existence of the shell $a$, the method would suggest that this shell should exist. In this case we would be left with some doubt, because another identification of shells $b$, $c$ and $d$ is possible for a~small fraction of potentials, which gives them shell numbers 2, 3, and 4.}
\label{fig:idtab_test}
\end{figure}

%%%%%%%%%%%%%%%%%%%%%%%%%%%%%%%%%%%%%%%%%%%%%%%%%%%%%%%%%%%%%%%%%
%%%%%%%%%%%%%%%%%%%%%%%%%%%%%%%%%%%%%%%%%%%%%%%%%%%%%%%%%%%%%%%%%
\subsection{Predictive power of the identification method} \label{sec:predictive_power}
To test the predictive ability of our method, we tried to ``rediscover'' the outermost already known shell $a$, which lies at 1170$\arcsec$ from the center of NGC\,3923. We excluded it from the set of shells that we feed into our identification code. We repeated the work described in Sects.~\ref{sec:grid} and \ref{sec:interpol} (two different choices of the interpolation function) with this reduced set of shell radii. The result of the identification for the basic grid of potentials is shown in Fig.~\ref{fig:idtab_test}, which is analogous to Fig.~\ref{fig:idtab_basgrid}.

The three outermost shells of this reduced set have $n=3$, 4, and 5 for all potentials compatible with the reduced set of shell radii of NGC\,3923 and for both interpolation functions, implying that two new shells must exist.   

If we assume that the correct interpolation function is the ``simple'' function, the predicted median radius of shell number 2 is located at +1168$\arcsec$ or +1210$\arcsec$ for the case of the ``standard'' function. The statistics of the predicted radius of shell number 2 are listed in Table~\ref{tab:possh2}. For the most probable free parameters of the potential (those from \citealp{bil13}), we derive +1149$\arcsec$ for the ``simple'' interpolation function and +1171$\arcsec$ for the ``standard'' interpolation function. If we had relied on the average value of these two most probable values, +1160$\arcsec$, we would have missed the correct value by only 1\%.

However, for a~small but not negligible part of the tested potentials (for both cases of interpolation functions), 
an alternative identification of the three shells ($b$, $c$, $d$) is possible. It assignes them  shell numbers 2, 3, and 4 and classifies them into the first generation. This makes the reconstruction of the radius of the excluded shell $a$ ambiguous. If we tried to predict the radius of shell 1 and shell $a$ had
been discovered, its radius would be substantially different from the predicted value. 

Nevertheless, we are in a better situation in reality -- we already know the radius of the shell $a$.  With this additional information the prediction of the position of the new shell is unambiguous.

\begin{table}
        \centering
\caption{Statistics of the predicted radius of shell number 2.}
                \begin{tabular}{lcccc}\hline\hline
        & $     r_{1,\mathrm{median}}$          & RMS           & $r_{1,\mathrm{min}}$    & $r_{1,\mathrm{max}}$\\\hline
basic grid  & 1169       & 21   & 1139 & 1224\\
``standard'' $\mu$       & 1210  & 28 & 1625 &  1254\\\hline
                \end{tabular}
                \tablefoot{``Rediscovering'' the outermost known shell $a$ at 1170$\arcsec$. All quantities are in arcseconds and their meaning is the same as in Table~\ref{tab:possh1}.}
                \label{tab:possh2}
\end{table}

%%%%%%%%%%%%%%%%%%%%%%%%%%%%%%%%%%%%%%%%%%%%%%%%%%%%%%%%%%%%%%%%%
%%%%%%%%%%%%%%%%%%%%%%%%%%%%%%%%%%%%%%%%%%%%%%%%%%%%%%%%%%%%%%%%%
\subsection{Newtonian self-consistent simulation} \label{sec:newton}
Many simplifying assumptions were made when deriving the analytical formulas (\ref{eq:pos1})\,--\,(\ref{eq:pos2}) to computate the shell radii. To test their precision, we applied our code to the results of a~Newtonian self-consistent simulation of a~shell galaxy formation using GADGET-2 \citep{springel05}. 
The simulation was an enhanced version of that presented in \citet{bartoskovaselfcon}. The primary galaxy was modeled as a~composite of two Plummer spheres corresponding to stellar and dark matter components. The secondary was modeled by a~single Plummer sphere. The collision was exactly radial. The specifications of the mass profiles and initial conditions are listed in Table~\ref{tab:gadget}.
Two generations of shells formed before the secondary was completely disrupted.

We made snapshots of the simulation at three different times. We measured the shell positions in each of them as we would in the image of a~real galaxy. Then we entered the shell radii into our identification code (along with the known Newtonian potential of the simulated primary). In this case, we had the privilege to know, in any snapshot, in how many generations the shells were formed (i.e., we knew $N_\mathrm{max}$, see Sect.~\ref{sec:aut_id}). The code indeed found the correct identification for each snapshot. 
In a~few cases, we were even led by the code to look at the images more carefully because it suggested that other shells should exist and they really showed up to be present, although faint. The correct identification was always found for the lowest possible value of the parameter $\Delta$ -- for lower values of $\Delta$ no satisfactory identification exists. In none of the three snapshots the relative difference of the modeled and measured shells exceeded 4\%.

We were able to see five shells in the snapshot in which only the first generation of shells was created, and seven and ten shells in the two snapshots in which two shell generations were already created. This simulation supports the claim that the one-generation scenario is unable to explain the high number of shells observed in some galaxies.

\begin{table}[t]
\centering
\caption{Parameters of the Newtonian self-consistent simulation of the~shell galaxy formation.}
\begin{tabular}{lccc}
\hline \hline 
component & $M$ & $b$ & $N_\mathrm{p}$ \\
 & $10^{10}$\,$M_{\sun}$ & kpc & $10^6$\\
\hline 
primary dark halo & 800 & 20 & 1.6\\
primary luminous matter & 20 & 8 & 0.4\\
secondary & 2 & 2 & 5\\
\hline \hline
init. primary-secondary separation & 200 & kpc & \\
init. relative velocity & 102 & km$/$s & \\
\hline 
\end{tabular}
\tablefoot{$M$ -- the total mass of a~Plummer sphere, $b$ -- Plummer radius, $N_\mathrm{p}$ -- number of particles}
\label{tab:gadget}
\end{table}

%%%%%%%%%%%%%%%%%%%%%%%%%%%%%%%%%%%%%%%%%%%%%%%%%%%%%%%%%%%%%%%%%
%%%%%%%%%%%%%%%%%%%%%%%%%%%%%%%%%%%%%%%%%%%%%%%%%%%%%%%%%%%%%%%%%
\subsection{Limitations of our method} \label{sec:unexpl}
The Newtonian self-consistent simulation gives us an idea, at least for one particular simulation, of the magnitude of influence of phenomena such as the noninstant disruption of the secondary, its nonzero size, and the self-gravity of its material. These phenomena were not taken into account when deriving the key relations Eqs.~(\ref{eq:pos1})\,--\,(\ref{eq:pos2}) for the time evolution of the shell radii. However, the simulation was still somewhat idealized and it was not MONDian. Here we try to list some of the remaining possible shortcomings of our model and of our conclusions.

1) \textit{Nonradiality of the merger and the ellipticity of the potential.}
The merger that produced the shells in NGC\,3923 was hardly exactly radial. The simulations of \citet{hq87} and \citet{DC86}
demonstrated the focusing effect of the host ellipticity on shells -- the higher the ellipticity of the potential, the narrower the angular extent of the shell system. The focusing effect probably also helps to create an axisymmetric shell system, even if the merger is not exactly radial. In this case, the orbits of stars would deviate more from the radial 
trajectories and Eqs.~(\ref{eq:pos1})\,--\,(\ref{eq:pos2}), which
are  the basis of our identification code, would lose precision.

2) \textit{Faster disruption of the secondary because of the external field effect (EFE).} 
It has been known for a long time that the galactic interactions in MOND are more complicated than in the Newtonian dynamics because
of the nonlinearity of MOND equations. In MOND, the dynamics of a~system is affected by the presence of an external field, even if the external field is homogeneous (i.e., the strong equivalence principle is violated). 
The process of satellite disruption in a~highly elliptical orbit was discussed in detail by \citet{satelliteefe}. The external field effect causes an expansion of the satellite, so that it becomes more vulnerable to tides. 
In the era when the one-generation shell formation scenario was accepted, \citet{milgsh} suggested that when a~shell galaxy is formed, the secondary becomes gradually disrupted sooner than it reaches the center of the primary because of the EFE. This would mean that two of the assumptions of Eq.~(\ref{eq:pos1}) are incorrect. The question is how important this effect was in the case of NGC\,3923. 
Since the radial range of shells is so large, 
the shell system probably had to be formed in several generations. 
Since the mass-loss of the secondary due to tidal forces is strongest near the center of the primary, most of the material that forms shells was probably released at the center, as our model expects. If much material was released before the secondary reached the center, then the radii of the shells made of this material could, indeed, deviate substantially from our model.

3) \textit{Influence of the galactic neighbors.}
The galaxy NGC\,3923 is the largest member of a~small galactic group. A~nearby small elliptical galaxy NGC\,3904 is seen ~2200$\arcsec$ northwest from NGC\,3923. This is very close to the expected position of the predicted shell. The distance of NGC\,3904 has been measured by the method of the surface brightness fluctuations. There are three individual measurements in the NASA/IPAC Extragalactic Database ranging from 28.3 to 29.8\,Mpc. This means that the galaxy is at least 5\,Mpc farther away from Earth than NGC\,3923. 
The apparent V magnitude is lower by 1 than that of NGC\,3923, therefore the stellar mass of NGC\,3904 is about 1.7\,$\times$\,lower than that of NGC\,3923, if their mass-to-light ratio are approximately equal. Therefore its MONDian acceleration is $3\times 10^{-3}a_0$ at a distance of 5\,Mpc. Thus the effect of NGC\,3904 on the position of the predicted shell is negligible because it is subject to acceleration of $0.1a_0$ from NGC\,3923.

4) \textit{Formation of shells from several minor mergers.} 
The shell system could, of course, be created by accretion of several secondaries, not only by a single one as our method assumes. However, this is highly improbable. Type~I shell galaxies constitute about 3\% of all early-type galaxies \citep{prieur90}. From this number, we can easily deduce that the probability that the shell system in  a~randomly chosen Type\,I shell galaxy was formed by accretion of $k$ independent secondaries is $33/34^k$, that
is, only 3\% of all Type\,I shell systems come from more than one secondary. Furthermore, the actual probability is even lower because the two or more independent shell systems would have to share a common symmetry axis when viewed from Earth. 

5) \textit{Other formation scenarios of shell galaxies exist.} 
We assumed a minor-merger origin of the shell system in NGC\,3923. However, other scenarios of shell formation also exist, but they are even less explored than the minor merger model. The merger scenario is favored in the case of NGC\,3923 because the galaxy contains nonthermalized dust clouds \citep{sikkema07}. 
The so-called weak interaction model of the formation of shell galaxies requires a~rotationally supported thick stellar disk \citep{wim90,wim91}. For a~Type\,I shell galaxy, the thick disk has to lie in the plane of symmetry of the shell system, but a minor axis rotation of NGC\,3923 is observed \citep{minrot}. This rules out the weak interaction model.
 \citet{majorm} claimed that the result of a~merger of two identical spiral galaxies formed in their simulation resembled NGC\,3923. In such a~merger, our model of shell radii evolution is not applicable.
 
6) \textit{What is the radius of a~shell?} In our simplified model, where the stars are all released simultaneously from a~single point, the shell edges are discontinuities in the surface density. In simulations, and even more so in observed galaxies, the shell edges are blurred. What should then be considered as the shell radius? The radius of the highest surface brightness of the shell, or the radius of its steepest gradient? These two radii can differ by several percent. For now, we included this observation uncertainty, jointly with the systematic errors caused by the inaccuracy of our model of shell propagation, into the tolerance parameter $\Delta$.
 
7) \textit{Arbitrariness of the identification criteria.} We deliberately chose a wide range of parameters to test whether the identification of the four outermost shells was unique. However, the criteria were still arbitrary, but it will be possible to set the criteria more correctly when more work has been done. We can learn about the tolerance parameter $\Delta$ from self-consistent simulations. These simulations will have to include nonradial mergers, elliptical primaries, different secondary morphologies, etc. The best measure of the deviation of the modeled from the observed shell radii may be different from the relative difference we used here. We can imagine that the best measure can be defined, for example, as $\left|r_{\lambda}-r_\mathrm{model}\right|<C$ or $\left|r_{\lambda}-r_\mathrm{model}\right|<C\sqrt{r_{\lambda}}$ instead of $\left|r_{\lambda}-r_\mathrm{model}\right|/\left|r_{\lambda}\right|<\Delta$. The self-consistent simulations will also indicate the largest possible number of shell generations present in the system. This number will probably be connected with the values of the gravitational potential at the positions of the innermost and outermost observable shells. The possible number of missing shells can be constrained from observations.

%%%%%%%%%%%%%%%%%%%%%%%%%%%%%%%%%%%%%%%%%%%%%%%%%%%%%%%%%%%%%%%%%
%%%%%%%%%%%%%%%%%%%%%%%%%%%%%%%%%%%%%%%%%%%%%%%%%%%%%%%%%%%%%%%%%
\section{Summary} \label{sec:summary}

\begin{figure}
\sidecaption
\resizebox{\hsize}{!}{\includegraphics{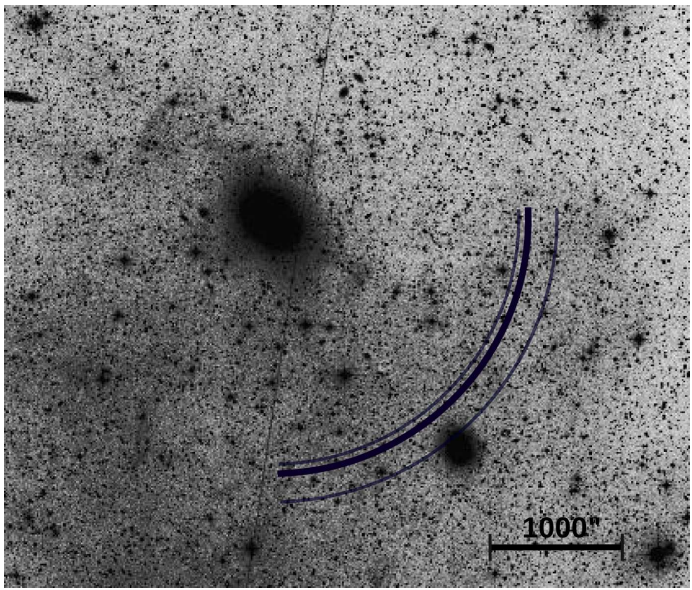}} % one column fig.
\caption{Position of the predicted shell in NGC\,3923.  The middle thick arc is the most probable position of the predicted shell (calculated for the mass distribution used in \citealp{bil13}). The inner and the outer arcs present the error range deduced as the lowest and highest values in  Table~\ref{tab:shpos} excluding the point mass potentials. The data for the underlying image are taken from the Digitized Sky Survey.}
\label{fig:pred}
\end{figure}

The elliptical galaxy NGC\,3923 is a well-studied case of a~shell galaxy with many exceptional properties. According to our method, it is necessary in MOND for the four outermost known shells (labels $a$, $b$, $c$, and $d$ in Table~\ref{tab:shpos}) to have shell numbers 2, 3, 4, and 5. This implies that shell number 1 must exist. 
The position of the predicted shell depends on the free parameters of the potential, see Table~\ref{tab:shpos}. Most probably, the shell lies in an angular distance of about 1900$\arcsec$ (Fig.~\ref{fig:pred}, Sect.~\ref{sec:resdis}) from the galaxy center to the southwest. The shell could be observable by existing instruments. In addition
to the shell, up to three stellar tails could be discovered -- two pointing at the northeastern side of the galaxy and one pointing at the southwestern side. 
We verified that this conclusion does not change when several uncertain assumptions are modified. 
We varied the mass-to-light ratio, the distance of the galaxy from Earth, the interpolation function, the value of the acceleration constant $a_0$, the density profile of the galaxy, and the treatment of a~problematic shell $i$. 

These results were achieved using our code for shell identification (identification means assigning two orbital quantities -- the shell and the generation number -- to each of the 25 outermost observed shells of NGC\,3923). 
The code tests whether the set of observed shell radii is consistent with a~given gravitational potential (that means that it is possible to find an identification of the shell set, so that the criteria described in Sect.~\ref{sec:aut_id} are satisfied). The identification criteria are forced by the minor-merger model of shell formation. The code uses analytical expressions to evolve the shell radii in time in a~given potential, Eq.~(\ref{eq:pos2}). The functionality of the code (and of the analytical equations) was successfully tested on a~self-consistent Newtonian simulation of a shell galaxy formation. In this simulation, the shell radii deviate from the analytical model by less than 4\%. 

To test the predictive ability of the method, we excluded the outermost known shell and tried to reconstruct its radius from the positions of the remaining shells. Using the same procedure as for the prediction of the new shell, we derived a radius of 1160$\arcsec$. The correct value is 1170$\arcsec$ (1\% deviation).

If the identification of the four outermost shells is correct, the accreted galaxy originally arrived at NGC\,3923 from the southwestern side. For the potentials from the basic grid (Sect.~\ref{sec:grid}), the age of the oldest shells in the system must be between 2 and 4\,Gyr.

We refrained from identifying shells with a radius smaller than 400$\arcsec$, since their identification is not unique. 
Their unique identification would be possible if we were sure that our model of the evolution of shell radii is sufficiently precise. This would require performing many MONDian self-consistent simulations. The prediction of the new outermost shell is valid provided that the radii of the real shells do not deviate from our analytical model by more than 10\%.

The discovery of the new shell at the predicted position would evidently support MOND. If no new shell is discovered, it would give no information about the validity of MOND because the surface brightness of the predicted shell might simply be below our detection limit. However, if a new shell is discovered at a substantially different position than we predict, or if two or more new shells are detected, it might be a problem for MOND. We will perform a similar analysis for the dark matter theory in a next paper. We expect that the prediction of the new shell will be more ambiguous than in MOND because the potential is not as tightly constrained in dark matter theory.

%-------------------------------------------------------------------------
\begin{acknowledgements}
We thank I.~Orlitov\'{a}, L.~J\'{i}lkov\'{a}, and L. Janekov\'{a} for useful comments on the manuscript.
The work of IE was partially supported by the EU grant GLORIA (FP-7 Capacities; No.~283783).
We acknowledge the support from the following sources: Czech support for the
long-term development of the research institution RVO67985815
(MB, KB, IE, and BJ), the project SVV-260089 by the Charles University
in Prague (MB, and IE) and the grant MUNI/A/0773/2013 by the Masaryk
University in Brno (KB). 
%METACENTRUM
Access to computing and storage facilities owned by parties and projects
contributing to the National Grid Infrastructure MetaCentrum, provided
under the program "Projects of Large Infrastructure for Research,
Development, and Innovations" (LM2010005), is greatly appreciated.
%ULOZISTE DAT
Access to the CERIT-SC computing and storage facilities provided under
the program Center CERIT Scientific Cloud, part of the Operational
Program Research and Development for Innovations, reg. no. CZ.
1.05/3.2.00/08.0144, is greatly appreciated.
%DSS
We acknowledge The Digitized Sky Survey \url{https://stdatu.stsci.edu/dss/acknowledging.html}.
\end{acknowledgements}
%----------------------------------------------------------------------------------

\bibliographystyle{aa}
\bibliography{citace}

\end{document}